\documentclass[12pt,twoside]{article}
\usepackage{amsmath}

\raggedright

\newcommand{\rd}[1]{\mathop{\mathrm{d}#1}}
\newcommand{\fract}[2]{{\textstyle\frac{#1}{#2}}}
\newcommand{\grad}{\vec\nabla}

\newcommand{\va}{\vec a}
\newcommand{\vA}{\vec A}

\newcommand{\bra}[1]{\bigl\langle #1 \bigr| }
\newcommand{\ket}[1]{\bigl| #1 \bigr\rangle}
\newcommand{\braket}[2]{\bigl\langle #1 \big|  #2 \bigr\rangle}

\newcommand{\numeq}[2]{\begin{equation}
#2
\label{#1}
\end{equation}}
\newcommand{\refeq}[1]{(\ref{#1})}

\let\vec\boldsymbol
\let\eps\varepsilon
\let\epsilon\varepsilon
\let\phi\varphi
\let\hat\widehat

\begin{document}
 
\title{Observations on noncommuting coordinates\\ and on fields depending on them}
\author{R. Jackiw\\
\small\it Center for Theoretical Physics\\ 
\small\it Massachusetts Institute of Technology\\ 
\small\it Cambridge, MA 02139-4307}

\date{\vspace*{-.25in}\footnotetext{TH-2002, Paris, France, July 2002\qquad
MIT-CTP\#3335}}

\maketitle

\abstract{\noindent
The original ideas about noncommuting coordinates are recalled. The connection between U(1)
gauge fields defined on noncommuting coordinates and fluid mechanics is explained.
\vspace*{-.25in}}


\pagestyle{myheadings}
\markboth{\small {\it R. Jackiw}}{\small  Observations on noncommuting coordinates and on
fields depending on them}
\thispagestyle{empty}


\section*{}

The idea that configuration-space coordinates may not commute
\numeq{e1}{
[x^i, x^j] = i\theta^{ij}
}
where $\theta^{ij}$ is a constant, anti-symmetric two-index object, has arisen recently from
string theory, but in fact it has a longer history.  Like many interesting quantum-mechanical
ideas, it was first suggested  by Heisenberg, in the late 1930s, who reasoned that coordinate
noncommutativity would entail a coordinate uncertainty and would ameliorate short-distance
singularities, which beset quantum fields. He told  his idea to Peierls, who eventually made use
of it when analyzing electronic systems in an external magnetic field, so strong that projection
to the lowest Landau level is justified.   But this phenomenological realization of
Heisenberg's idea did not address  issues in fundamental science, so Peierls told Pauli about
it, who in turn told Oppenheimer, who  asked his student Snyder to work it out and this  led to
the first published paper on the subject~\cite{r1}. 

The coordinate noncommutativity in the lowest Landau level is very similar to today's
string-theory origins of noncommutativity -- both rely on the presence of a strong
background field. Also, thus far, it is the only physically realized example of noncommuting
coordinates, so let me describe it in a little detail~\cite{r2}.
We consider the motion of a charged ($e$) and massive ($m$) particle in a constant magnetic
field ($B$) pointing along the $z$~direction.  All interesting physics is in the $x$--$y$ plane. The
Lagrangian for this planar motion is
\numeq{n2}{
L = \fract12 m(\dot x^2 +\dot y^2) + \frac ec (\dot x A_x + \dot y A_y) - V(x,y) 
}
where the vector potential $\vA$ can be chosen as $(0, xB)$ and $V(x,y)$ describes additional
interactions (``impurities''). In the absence of $V$, the quantum spectrum consists of the
well-known Landau levels $\ket{N,d}$, where $N$ indexes the level's energy eigenvalue,
and~$d$ describes the infinite degeneracy of each level. The separation between levels is
$O(B/m)$, so that in the strong magnetic field limit only the lowest Landau level $\ket{0,d}$ is
relevant. But observe that the large $B$ limit corresponds to small~$m$, so projection on the
lowest Landau level is also achieved by setting $m$ to zero in \refeq{n2}. In that limit the
Lagrangian \refeq{n2}, in the chosen gauge, becomes
\numeq{n3}{
L_{\ell L\ell} = \frac ec Bx\dot y - V(x,y)\ . 
}
This is of the form $p\dot q - H(p,q)$, and immediately identifies $\frac ec Bx$ and $y$ as
canonical conjugates, leading in the usual way to the commutator
\numeq{n4}{
[x,y] = - i \frac{\hbar c}{eB}\ .
}
[The ``Peierls substitution'' consists of determining the effect of the impurity by computing the
eigenvalues of $V(x,y)$, where $x$ and $y$ are noncommuting.]

For another perspective, consider a calculation of the lowest Landau level  matrix elements of
the $[x,y]$ commutator. 
\numeq{n5}{
\bra{0,d} xy - yx \ket{0,d'} = M(d,d') - M^* (d',d)
}
where
\numeq{n6}{
 M(d,d') = \bra{0,d} xy \ket{0,d'} \ .
}
We evaluate \refeq{n6} by inserting intermediate states in product $xy$:
\numeq{n7}{
 M(d,d') = \sum_s \bra{0,d} x  \ket{s} \bra{s}  y \ket{0,d'}\ . 
}
If the sum is over \emph{all} the degenerate Landau levels, then one finds that \refeq{n5}
vanishes: $x$ and $y$ do commute! But if one pretends that the world is restricted to the
lowest Landau level and includes only that level (with its degeneracy) in the intermediate
state sum 
\numeq{n8}{
M_{\ell L\ell} = \sum_{d''} \bra{0,d} x  \ket{0,d''} \bra{0,d''}  y \ket{0,d'}
}
one finds that in this truncated state space, eq.~\refeq{n7} becomes consistent with \refeq{n4}:
\numeq{n9}{
\bra{0,d} [x,y] \ket{0,d'} = -i \frac{\hbar c}{eB} \braket{0,d}{0,d'} \ . 
}

Let me now return to the general and abstract problem of noncommuting coordinates. 
When confronting the noncommutativity postulate~\refeq{e1}, it is natural to ask which
(infinitesimal) coordinate transformations 
\numeq{e2}{
\delta x^i = -f^i (x) 
}
leave \refeq{e1} unchanged. 
The answer is that the (infinitesimal) transformation vector function $f^i(x)$ must be
determined by a scalar $f(x)$ through the expression~\cite{r3}
\numeq{e3}{
f^i (x) = \theta^{ij} \partial_j f(x) \ . 
}
Since $\partial_i f^i(x) = 0$,   these are recognized as volume-preserving transformations.
[They do not exhaust all volume preserving transformations, except in two dimensions. In
dimensions greater  two,
\refeq{e3} defines a subgroup of volume-preserving transforms that also leave $\theta^{ij}$
invariant.]

The volume-preserving transformations form the link between noncommuting coordinates
and fluid mechanics. Since the theory of fluid mechanics is not widely known outside the circle
of fluid mechanicians, let me put down some relevant facts \cite{r4}.  There are two, physically
equivalent descriptions of fluid motion: One is the Lagrange formulation, wherein the fluid
elements are labeled, first by a discrete index~$n$: $\vec X_n(t)$ is the position as a function of
time of the $n$th fluid element.  Then one passes to a continuous labeling variable $n \to \vec
x: \vec X_n(t)\to \vec X (t,\vec  x)$, and 
$\vec x$ may be taken to be the position of the fluid element at initial time $\vec X(0,\vec x) =
\vec x$. This is a comoving description. Because labels can be arbitrarily rearranged, without
affecting physical content, the continuum description is invariant against volume-preserving
transformations of $\vec x$, and in particular, it is invariant against the specific
volume-preserving transformations
\refeq{e3}, provided the fluid coordinate $\vec X$ transforms as a scalar:
\numeq{e4}{
\delta_f  \vec X  = f^i (\vec x) \frac\partial{\partial x^i} \vec X = \theta^{ij} \partial_i \vec
X\partial_j f\ . 
 }

The common invariance of Lagrange fluids and of noncommuting coordinates is a strong hint of
a connection between the two.

Formula \refeq{e4} will take  a very suggestive form when we rewrite it in terms of a bracket
defined for functions of $\vec x$ by 
\numeq{e5}{
\bigl\{ \mathcal O_1(\vec x), \mathcal O_2(\vec x)\bigr\} =\theta^{ij} \partial_i\mathcal
O_1(\vec x)
\partial_j\mathcal O_2(\vec x)\ .
} 
Note that with this bracket we have
\numeq{e6}{
\bigl\{x^i,  x^j\bigr\} = \theta^{ij}\ .
} 
So we can think of bracket relations as classical precursors of commutators for a
noncommutative field theory -- the latter obtained from the former by replacing brackets by
$-i$~times commutators, \`a la Dirac. More specifically, we shall see that the noncommuting field
theory that emerges from the Lagrange fluid is a noncommuting U(1) gauge theory. 

This happens when the following steps are taken. We define the evolving portion of $\vec X$ by 
\numeq{e7}{
X^i (t,\vec x) = x^i + \theta^{ij} \hat A_j (t,\vec x)\ .
}
(It is assumed that $\theta^{ij}$ has an inverse.)
Then \refeq{e4} is equivalent to the suggestive expression
\numeq{e8}{
\delta_f \hat A_i = \partial_i f + \bigl\{\hat A_i, f\bigr\}\ .
}
When the bracket is replaced by $(-i)$ times the commutator, this is precisely the gauge
transformation for a noncommuting U(1) gauge potential $\hat A_i$. Moreover, the gauge field
$\hat F_{ij}$ emerges from the bracket of two Lagrange coordinates
\begin{gather}
\bigl\{ X^i, X^j\bigr\} = \theta^{ij} + \theta^{im} \theta^{jn} \hat F_{mn} \label{e9}\\
\hat F_{mn}  = \partial_m \hat A_n - \partial_n \hat A_m + \bigl\{\hat A_m, \hat  A_n\bigr\}\ . 
\label{e10}
\end{gather}
Again \refeq{e10} is recognized from the analogous formula
in noncommuting gauge theory.

What can one learn from the parallelism of the formalism for a Lagrange fluid and a
noncommuting gauge field? One result that has been obtained addresses the question of what
is  a gauge field's covariant response to a coordinate transformation. This question can be put
already for commuting, non-Abelian gauge fields, where conventionally the response is
given in terms of a Lie derivative $L_f$:
\begin{gather}
\delta_f x^\mu = - f^\mu(x) \label{e11}\\
\delta_f A_\mu = L_f A_\mu \equiv f^\alpha \partial_\alpha A_\mu + \partial_\mu f^\alpha
A_\alpha
 \ . 
\label{e12}
\end{gather}
But this implies
\numeq{e13}{
\delta_f F_{\mu\nu} = L_f F_{\mu\nu} \equiv f^\alpha \partial_\alpha F_{\mu\nu} + 
\partial_\mu f^\alpha F_{\alpha\nu} +  \partial_\nu f^\alpha F_{\mu\alpha}  
}
which is not covariant since the derivative in the first term on the right is not the covariant
one. The cure in this, commuting, situation has been given some time ago~\cite{r5}:  Observe
that
\refeq{e12} may be equivalently presented as 
\numeq{e14}{
\begin{aligned}
\delta_f A_\mu = L_f A_\mu &= f^\alpha \bigl(
\partial_\alpha A_\mu -\partial_\mu A_\alpha - i [A_\alpha, A_\mu]
\bigr) \\
&\qquad{}+ f^\alpha \partial_\mu A_\alpha  - i [  A_\mu, f^\alpha A_\alpha] + 
\partial_\mu f^\alpha A_\alpha\\
 &= f^\alpha F_{\alpha\mu} +  D_\mu (f^\alpha A_\alpha)\ .
\end{aligned}
}
Thus, if the coordinate transformation  generated by $f^\alpha$ is supplemented by a  gauge
transformation generated by $-f^\alpha A_\alpha$, the result is a gauge covariant coordinate
transformation
\numeq{e15}{
\delta'_f A_\mu = f^\alpha F_{\alpha\mu}
}
and the modified response of $F_{\mu\nu}$ involves the gauge-covariant Lie derivative $L'_f$:
\numeq{e16}{
\delta'_f F_{\mu\nu} = L'_f F_{\mu\nu} \equiv f^\alpha D_\alpha F_{\mu\nu} + 
\partial_\mu  f^\alpha F_{\alpha\nu} +  \partial_\nu f^\alpha  F_{\mu\alpha} \ .
}

In the noncommuting situation, loss of covariance in the ordinary Lie derivative is even
greater, because in general the coordinate transformation functions $f^\alpha$ do not
commute with the fields $A_\mu, F_{\mu\nu}$; moreover, multiplication of $x$-dependent
quantities is not a covariant operation. All these issues can be addressed and resolved by
considering them in the fluid mechanical context, at least, for linear and volume-preserving
diffeomorphisms. The analysis is technical and I refer you to the published papers~\cite{r3,r6}.
The final result for the covariant coordinate transformation on the noncommuting gauge
potential $\hat A_\mu$, generated by $f^\alpha (X)$, is 
\numeq{en25}{
\delta'_f \hat A_\mu = \fract12 \bigl\{
f^\alpha(X), \hat F_{\alpha\mu}\bigr\} + \mbox{reordering terms.}
}
Note that the generating function $f^\alpha(X)$ enters the anticommutator with covariant
argument~$X$. $f^\alpha$ is restricted to be either linear or volume-preserving; in the latter
case there are reordering terms, whose form is explicitly determined by the fluid mechanical
antecedent.

Next, I shall discuss the Seiberg-Witten map~\cite{r7}, which can be made very
transparent by the fluid analogy.  The Seiberg-Witten map replaces the noncommuting vector
potential $\hat A_\mu$  by a nonlocal function of a commuting potential $a_\mu$ and
of~$\theta$;  i.e., the former is viewed as a function of the latter.  The relationship between the
two follows from the requirement of stability against gauge transformations: a noncommuting
gauge transformation  of the noncommuting gauge potential should be equivalent to a
commuting gauge transformation on the commuting vector potential on which the
noncommuting potential depends. Formally:
\numeq{en26}{
\hat A_\mu(a + \rd \lambda) = \hat A_\mu^{G(a,\lambda)} (a)\ .
}
Here $\lambda$ is the Abelian gauge transformation function that transforms the Abelian,
commuting gauge potential $a_\mu$; $G(a,\lambda)$ is the noncommuting gauge function that
transforms the noncommuting gauge potential $\hat A_\mu$. $G$ depends on $a_\mu$
and~$\lambda$, and one can show that it is a noncommuting 1-cocycle~\cite{r8n}.

Moreover, when the action and the equations of motion of
the noncommuting theory are transformed into commuting variables, the dynamical content
is preserved: the physics described by noncommuting variables is equivalently described by
the commuting variables, albeit in a complicated, nonlocal fashion. 

  The Seiberg-Witten map is intrinsically interesting in the unexpected equivalence that it
establishes. Moreover, it is practically useful for the following reason. It is difficult to extract
gauge invariant content from a noncommuting gauge theory because quantities constructed
locally from $\hat F_{\mu\nu}$ are not gauge invariant; to achieve gauge invariance, one must
integrate over space-time. Yet for physical analysis one wants local quantities: profiles of
propagating waves, etc.  Such local quantities can be extracted in a gauge invariant manner
from the physically equivalent, Seiberg-Witten mapped commutative gauge theory~\cite{r8}. 

Let me now use the fluid analogy to obtain an explicit formula for the Seiberg-Witten map;
actually, we shall present the inverse map, expressing commuting fields in terms of
noncommuting ones. For our development we must refer to a second, alternative formulation
of fluid mechanics, the so-called Euler formulation. This is not a comoving description, rather
the experimenter observes the fluid density~$\rho$ and velocity~$\vec v$ at given point in
space-time $(t,\vec r)$. The current is $\rho \vec v$ and satisfies with $\rho$ a continuity
equation
\numeq{e17}{
\frac\partial{\partial t} \rho + \grad \cdot (\rho\vec v) = 0\ . 
}
The theory is completed by  positing an ``Euler equation'' for $\partial\vec v/\partial t$, but
we shall not need this here. 

Of interest to us is the relation between the Lagrange description and the Euler description.
This is given by the formulas
\begin{subequations}\label{e18}
\begin{gather}
\rho(t,\vec r) = \int \rd x \delta\bigl(\vec X(t,\vec x) -\vec r\bigr)\label{e18a}\\
\rho(t,\vec r)\vec v(t,\vec r) \equiv \vec j(t,\vec r) = 
    \int \rd x \frac\partial{\partial t} \vec X(t,\vec x) \delta \bigl(\vec X(t,\vec x) -\vec
r\bigr)\ . \label{e18b}
\end{gather}
\end{subequations}
(The integration and the $\delta$-function carry the dimensionality of space.) Observe that
the continuity equation \refeq{e17} follows from the definitions \refeq{e18}, which can be
summarized by 
\begin{gather}
j^\mu(t,\vec r) = \int \rd r \frac{\partial}{\partial t}  X^\mu \delta (\vec X - \vec
r)\label{e19}\\ X^0 = t\notag\\
\partial_\mu  j^\mu = 0\ . \label{e20}
\end{gather}

The (inverse) Seiberg-Witten map, for the case of two spatial dimensions, can be extracted
from \refeq{e19}, \refeq{e20}~\cite{r3}.  (The argument can be generalized to arbitrary
dimensions, but there it is more complicated~\cite{r3}.) Observe that the right side of
\refeq{e19} depends on
$\hat {\vA}$ through
$\vec X$ [see
\refeq{e7}]. It is easy to check that the integral \refeq{e19} is invariant under the
transformations \refeq{e4}; equivalently viewed as a function of $\hat {\vA}$, it is gauge
invariant   [see \refeq{e8}]. Owing to the conservation of $j^\mu$   [see \refeq{e20}], its dual 
$\eps_{\alpha\beta\mu} j^\mu$ satisfies a conventional, commuting Bianchi identity, and
therefore can be written as the curl of an Abelian vector potential~$a_\alpha$, apart from
proportionality  and additive constants: 
\numeq{e21}{
\begin{gathered}
\partial_\alpha a_\beta - \partial _\beta a_\alpha + \text{constant} 
\propto \eps_{\alpha\beta\mu} \int \rd x \frac\partial{\partial t} X^\mu \delta(\vec X- \vec
r)\\
\partial_i a_j - \partial _j a_i + \text{constant} 
\propto \eps_{ij} \int \rd x   \delta(\vec X- \vec r) = \eps_{ij} \rho\ .
\end{gathered}
 }
This is the (inverse) Seiberg-Witten map, relating the~$\va$ to~$\hat {\vA}$. 

Thus far operator noncommutativity has not been taken into account. To do so, we must
provide an ordering for the $\delta$-function depending on the operator $X^i = x^i + \theta^{ij}
\hat A_j$.  This we do with the Weyl prescription by Fourier transforming. The  final operator
version of equation~\refeq{e21}, restricted to the two-dimensional spatial components, reads
\numeq{e22}{
\int \rd r e^{i\vec k\cdot\vec r} (\partial_i a_j - \partial_j a_i) = 
-\eps^{ij} \Bigl[ \int \rd x  e^{i\vec k\cdot\vec X} - (2\pi)^2\delta(\vec k)   \Bigr] 
}
where the $x$ integral over an operator ($\vec X$) dependent integrand is interpreted as a
trace.  Here the additive and proportionality constants are determined by requiring
agreement for weak noncommuting fields.

Formula~\refeq{e22} has previously appeared  in a direct analysis of the Seiberg-Witten
relation~\cite{r9}. Now we recognize it as the (quantized) expression relating Lagrange and
Euler formulations for fluid mechanics. 

 \newpage 
\def\Journal#1#2#3#4{{\em #1} {\bf #2}, #3 (#4)}
\def\add#1#2#3{{\bf #1}, #2 (#3)}
\def\Book#1#2#3#4{{\em #1}  (#2, #3 #4)}
\def\Bookeds#1#2#3#4#5{{\em #1}, #2  (#3, #4 #5)}

\end{document}